\documentclass[aps,prc,twocolumn,amsfonts,amsmath,amssymb,showpacs,superscriptaddress,floatfix]{revtex4-1}

\usepackage{graphicx}
\usepackage{longtable}
\usepackage{bm}
\usepackage{multirow}
\usepackage{hyperref}
\usepackage{ulem}
\hypersetup{colorlinks,citecolor=blue,filecolor=blue,linkcolor=blue,urlcolor=blue}
\usepackage{color}
\usepackage{dcolumn}

\begin{document}

\title{Change of nuclear configurations in the neutrinoless double-{\boldmath$\beta$} decay of\\ {\boldmath$^{130}$}Te~{\boldmath$\rightarrow$~$^{130}$}Xe and {\boldmath$^{136}$}Xe~{\boldmath$\rightarrow$~$^{136}$}Ba}

\author{J.~P.~Entwisle}
\affiliation{School of Physics and Astronomy, University of Manchester, Manchester M13 9PL, United Kingdom}
\author{B.~P.~Kay}
\email[E-mail: ]{kay@anl.gov}
\affiliation{Physics Division, Argonne National Laboratory, Argonne, Illinois 60439, USA}
\author{A.~Tamii}
\affiliation{Research Center for Nuclear Physics (RCNP), Osaka University, Ibaraki, Osaka 567-0047, Japan}
\author{S.~Adachi}
\affiliation{Research Center for Nuclear Physics (RCNP), Osaka University, Ibaraki, Osaka 567-0047, Japan}
\author{N.~Aoi}
\affiliation{Research Center for Nuclear Physics (RCNP), Osaka University, Ibaraki, Osaka 567-0047, Japan}
\author{J.~A.~Clark}
\affiliation{Physics Division, Argonne National Laboratory, Argonne, Illinois 60439, USA}
\author{S.~J.~Freeman}
\affiliation{School of Physics and Astronomy, University of Manchester, Manchester M13 9PL, United Kingdom}
\author{H.~Fujita}
\affiliation{Research Center for Nuclear Physics (RCNP), Osaka University, Ibaraki, Osaka 567-0047, Japan}
\author{Y.~Fujita}
\affiliation{Research Center for Nuclear Physics (RCNP), Osaka University, Ibaraki, Osaka 567-0047, Japan}
\author{T.~Furuno}
\affiliation{Department of Physics, Kyoto University, Kyoto, 606-8502, Japan}
\author{T.~Hashimoto}
\affiliation{Rare Isotope Project, Institute for Basic Science, 70, Yuseong-daero, 1689-gil, Yuseong-gu, Daejeon, Korea}
\author{C.~R.~Hoffman}
\affiliation{Physics Division, Argonne National Laboratory, Argonne, Illinois 60439, USA}
\author{E.~Ideguchi}
\affiliation{Research Center for Nuclear Physics (RCNP), Osaka University, Ibaraki, Osaka 567-0047, Japan}
\author{T.~Ito}
\affiliation{Research Center for Nuclear Physics (RCNP), Osaka University, Ibaraki, Osaka 567-0047, Japan}
\author{C.~Iwamoto}
\affiliation{Research Center for Nuclear Physics (RCNP), Osaka University, Ibaraki, Osaka 567-0047, Japan}
\author{T.~Kawabata}
\affiliation{Department of Physics, Kyoto University, Kyoto, 606-8502, Japan}
\author{B. Liu}
\affiliation{Research Center for Nuclear Physics (RCNP), Osaka University, Ibaraki, Osaka 567-0047, Japan}
\author{M.~Miura}
\affiliation{Research Center for Nuclear Physics (RCNP), Osaka University, Ibaraki, Osaka 567-0047, Japan}
\author{H.~J.~Ong}
\affiliation{Research Center for Nuclear Physics (RCNP), Osaka University, Ibaraki, Osaka 567-0047, Japan}
\author{J.~P.~Schiffer}
\affiliation{Physics Division, Argonne National Laboratory, Argonne, Illinois 60439, USA}
\author{D.~K.~Sharp}
\affiliation{School of Physics and Astronomy, University of Manchester, Manchester M13 9PL, United Kingdom}
\author{G.~S\"{u}soy}
\affiliation{Physics Department, Faculty of Science, Istanbul University, 34459 Vezneciler, Istanbul, Turkey}
\author{T.~Suzuki}
\affiliation{Research Center for Nuclear Physics (RCNP), Osaka University, Ibaraki, Osaka 567-0047, Japan}
\author{S.~V.~Szwec}
\affiliation{School of Physics and Astronomy, University of Manchester, Manchester M13 9PL, United Kingdom}
\author{M.~Takaki}
\affiliation{Center for Nuclear Study (CNS), University of Tokyo, Bunkyo, Tokyo 113-0033, Japan}
\author{M.~Tsumura}
\affiliation{Department of Physics, Kyoto University, Kyoto, 606-8502, Japan}
\author{T.~Yamamoto}
\affiliation{Research Center for Nuclear Physics (RCNP), Osaka University, Ibaraki, Osaka 567-0047, Japan}

\date{\today}

\begin{abstract}

The change in the configuration of valence protons between the initial and final states in the neutrinoless double-$\beta$ decay of $^{130}$Te~$\rightarrow$~$^{130}$Xe and of $^{136}$Xe~$\rightarrow$~$^{136}$Ba has been determined by measuring the cross sections of the ($d$,$^3$He) reaction with 101-MeV deuterons. Together with our recent determination of the relevant neutron configurations involved in the process, a quantitative comparison with the latest shell-model and interacting-boson-model calculations reveals significant discrepancies. These are the same calculations used to determine the nuclear matrix elements governing the rate of neutrinoless double-$\beta$ decay in these systems.

\end{abstract}


\maketitle

\section{Introduction} 

The prospect of observing neutrinoless double-$\beta$ ($0\nu2\beta$) decay is of great current interest, and is considered an essential probe to address outstanding questions concerning the nature of the neutrino~\cite{avignone,rodejohan,gomez,vergados}. Its observation would immediately inform us that the neutrino is Majorana in nature and thus, that lepton number is not a conserved quantity, demanding modifications to the standard model. Beyond that, a measurement of the half-life of $0\nu2\beta$ decay would provide access to the effective mass of the neutrino and thus a scale for the absolute mass of the neutrino. This requires knowledge of the nuclear matrix elements for this process.

The nuclear matrix elements for $0\nu2\beta$ decay are based on theoretical calculations using various nuclear-structure models. For any given $0\nu2\beta$-decay candidate the results vary by a factor of 2-3, which translates to as much as an order of magnitude in the predicted half-lives.

Reducing this discrepancy in the nuclear matrix elements is a major challenge. It is still not clear which theoretical approach is most applicable and what ingredients are most relevant (see, for example, Ref.~\cite{vergados}). Certain experimental data can provide important constraints on such calculations. There is no simple connection between double-$\beta$ decay with and without neutrinos, as far as the nuclear matrix element is concerned~\cite{vergados}. The change in the ground-state nucleon occupancies {\it must} be important~\cite{schifferge,freeman}. Which neutrons decay and which protons are created in the decay, and how their configurations are rearranged, can be probed in single-nucleon transfer reactions to a level of precision corresponding to a few tenths of a nucleon. Calculations can then be directly compared to the experimentally derived single-nucleon occupancies. The neutron and proton occupancies of the ground states of $^{76}$Ge and $^{76}$Se, and their change in the $^{76}$Ge~$\rightarrow$~$^{76}$Se decay, were published in Refs.~\cite{schifferge,kayge}. For that system, theoretical calculations explored the impact on the magnitude of the calculated nuclear matrix elements based on modifications to reproduce the experimental occupancies~\cite{sim1,sim2,suhonen_ge,menendezge} and found almost a factor of 2 reduction in the discrepancy between different models. 

There are several other candidate nuclei for large-scale experiments in search of $0\nu2\beta$ decay. Among these are  $^{130}$Te and $^{136}$Xe. $^{130}$Te is the isotope used in the CUORE experiment, which recently published half-life limits from its first stage CUORE-0 experiment~\cite{cuore}. Other searches include the COBRA experiment, which also recently published~\cite{cobra} a limit, and a  future experiment, SNO+~\cite{sno}, is under way. $^{130}$Te is favorable in terms of its high natural abundance, 34\%, and a moderately high $Q$ value of 2527~keV~\cite{redshaw,scielzo}. It has a long $2\nu2\beta$-decay half-life of $T^{2\nu}_{1/2}=7.0\times10^{20}$~yr~\cite{nemo3}, one of the longest of all candidates. A long $2\nu2\beta$-decay half-life is advantageous as it results in fewer background counts from this decay mode in the region one would expect the $0\nu2\beta$-decay peak. The best current limit of $T^{0\nu}_{1/2}$ for $^{130}$Te is provided by a combined analysis of Cuoricino and CUORE-0 data at $T^{0\nu}_{1/2}>4\times10^{24}$~yr~\cite{cuore}. 

$^{136}$Xe is the isotope used in the EXO(-200) and KamLAND-Zen experiments, recently reporting new limits of $T^{0\nu}_{1/2}>1.1\times10^{25}$~yr~\cite{exo} and $>1.9\times10^{25}$~yr~\cite{kamland}, respectively. It has the advantage of having the longest $T^{2\nu}_{1/2}$ of all practical candidates at $2\times10^{21}$~yr~\cite{exo2v}, a moderately high $Q$ value of 2458~keV~\cite{redshawxe}, and a natural abundance of 8.86\%.

In this paper, we report the change in the proton configurations between the parent and daughter in the  $^{130}$Te~$\rightarrow$~$^{130}$Xe and $^{136}$Xe~$\rightarrow$~$^{136}$Ba decays. As for the $^{76}$Ge~$\rightarrow$~$^{76}$Se system, in these nuclei both valence neutrons and protons are in the same major oscillator shell, lying relatively close to $Z=50$ and  $N=82$. This somewhat simplifies the nuclear structure, making them more conducive to shell-model studies. A previous publication has reported on the neutron occupancies for the $^{130}$Te~$\rightarrow$~$^{130}$Xe system~\cite{kayxe}, and results are forthcoming on the neutron occupancies for the $^{136}$Xe~$\rightarrow$~$^{136}$Ba system~\cite{kayxeba}. 

To determine the proton occupancies in these systems, we carried out a measurement of the ($d$,$^3$He) reaction on $^{130}$Te, $^{130,136}$Xe, and $^{136}$Ba in a consistent manner; additional measurements on the neighboring isotones---$^{128}$Te, $^{132,134}$Xe, and $^{138}$Ba---provided important checks.

The ($d$,$^3$He) reaction has been studied before on $^{128}$Te and $^{130}$Te at 34~MeV~\cite{auble_dh} and additionally the ($t$,$\alpha$) reaction was studied at 12~MeV~\cite{conjeaud} and later at 18~MeV~\cite{shahabuddin}. In all instances, there was a strong transition to the 7/2$^+$ ground state, and two weaker states, below 1~MeV, carried $\ell=2$ strength. There were also reports of a weak $\ell=0$ transition around 1--1.5~MeV in most cases. The work by Auble {\it et al.}~\cite{auble_dh} is the only one to publish cross sections. 

No reports of proton removal or addition reactions on Xe isotopes have been published with the exception of $^{136}$Xe. Cross sections and proton occupancies from the ($^3$He,$d$) and ($d$,$^3$He) reactions on the stable, even $N=82$ isotones, including $^{136}$Xe, were reported by Wildenthal {\it et al.}~\cite{wildenthal}, though the uncertainties were large, particularly for the $^{136}$Xe target. The same work reported results on proton adding and removing on $^{138}$Ba. No published data are available for proton transfer on $^{136}$Ba. Proton adding via both the ($^3$He,$d$)~\cite{auble_hd,szanto} and ($\alpha$,$t$)~\cite{szanto} reactions, have been carried out on $^{128,130}$Te leading to proton states in $^{129,131}$I. We take advantage of this complementary information in confirming spin assignments as these are, in most cases, the same final states populated in $^{130,132}$Xe($d$,$^3$He). 

\section{The experiment} \label{exp}

The experiment was carried out at the Research Center for Nuclear Physics (RCNP) at Osaka University, Japan. The coupling of the AVF and Ring Cyclotrons provided a 101-MeV beam of deuterons, which was delivered to the scattering chamber of the Grand Raiden (GR) spectrometer~\cite{gr} via the WS beam line. The dispersion matching capabilities were not used in this experiment. 

The ($d$,$^3$He) reaction was carried out on targets of $^{128,130}$Te, $^{130,132,134,136}$Xe, and $^{136,138}$Ba. The solid targets, made from isotopically-enriched materials, were of nominal thicknesses between $\sim$400-500~$\mu$g/cm$^2$ supported on carbon foils of thickness 100~$\mu$g/cm$^2$.  For the Xe isotopes, the Grand Raiden gas-target system was used~\cite{matsubara}. The target was of depth (along the beam line axis) of $\sim$8 mm. The windows were polyethylene naphthalate (PEN) foils~\cite{PEN} of thicknesses of 4 and 6~$\mu$m. Five gas cells were prepared in case of breakages during the experiment; however, only one was used with a window thickness of 4~$\mu$m. It lasted the duration of the measurement without any evidence of degradation (which can be assessed from the reactions on carbon and oxygen in the window). The windows withstood a total dose of $\sim$2$\times$10$^{16}$ deuterons at an average current of 20-30~nA over $\sim$37 hours. The beam spot was $\lesssim$2~mm in diameter. PEN contains only carbon, oxygen, and hydrogen. Reactions on carbon and oxygen result in manageable contaminants in the outgoing $^3$He spectra, appearing at excitation energies higher than the region of interest. Other plastics, which often contain nitrogen and chlorine, would result in peaks in the region of interest. The PEN foil windows had been used in a study of the $^{136}$Xe($^3$He,$t$)$^{136}$Cs reaction by Puppe {\it et al.}~\cite{puppe}. Using the empirical expression in Ref.~\cite{matsubara}, the average thickness was calculated to be $\sim$500~$\mu$g/cm$^2$. This was confirmed using elastically scattered deuteron yields. The pressure and temperature of the gas volume were monitored throughout the experiment, as discussed below.

The GR spectrometer was used to momentum analyze the outgoing ions. Vertical drift chambers and scintillators at the focal plane~\cite{gr} were used to record their positions and select the $^3$He ions. For the ($d$,$^3$He) measurements, the aperture was 1.36~msr, corresponding to an angular width of approximately $\pm$0.8$^{\circ}$. In order to estimate the absolute cross-section scale and to provide reliable relative cross sections between each of the targets, deuteron elastic scattering was carried out at the same incident beam energy as the ($d$,$^3$He) reaction. Typically a low-energy scattering measurement would be used at angles such that the cross section could be reliably assumed to be Rutherford scattering. With the gas target such a measurement is not possible, as the scattered ions would have insufficient energy to pass through the gas volume and the windows.  The elastic deuteron-scattering cross section was explored with four different optical-model calculations using different global parametrizations~\cite{an,han,daehnick,bojowald}. The calculated cross sections at the local maximum of $\theta_{\rm lab}=11.4^{\circ}$ varied by less than 8\% for a given isotope with different parametrizations, while the relative cross sections, from one nucleus to another, varied by less than 2\%. The fact that $\theta_{\rm lab}=11.4^{\circ}$ was indeed a local maximum in cross section, with a width approximately $\pm$1$^{\circ}$, was confirmed in measurements of the ($d$,$d$) reaction at three angles ($\theta_{\rm lab}=10.6^{\circ}$, 11.4$^{\circ}$, and 12.2$^{\circ}$) on all targets. For the ($d$,$d$) measurements a smaller aperture of 0.68~msr, corresponding to an angular width of $\pm$0.4$^{\circ}$, was used. Typical beam intensities were $\sim$30~nA for the ($d$,$^3$He) reaction and $\sim$1~nA for the ($d$,$d$) reaction.

The gases were isotopically enriched to greater than 99.9\%. The pressure and temperature of the loaded gas cells were monitored throughout the measurement, during periods both with and without beam. Variations were less than a few percent in pressure and temperature throughout the run. In addition to this continuous monitoring of pressure and temperature, the ($d$,$d$) reaction was measured before and after the longer ($d$,$^3$He)-reaction runs and normalized to the integrated beam current; these normalized yields from before and after each run were consistent at a level of $<$2\%, showing that the effective target thickness is constant to this level.

For the $^{130}$Te target, the ($d$,$^3$He) reaction was measured at six angles of $\theta_{\rm lab}=2.5^{\circ}$, 5.8$^{\circ}$, 9.0$^{\circ}$, 12.2$^{\circ}$, 15.4$^{\circ}$, and 18.0$^{\circ}$. Using the resulting angular distributions, an assessment could be made as to the suitability of different optical-model-potential parametrizations used in the distorted-wave Born approximation (DWBA) calculations. For all other targets, three angles were measured. These were $\theta_{\rm lab}=2.5^{\circ}$, 5.8$^{\circ}$, and 18$^{\circ}$. The two most forward angles are close to the first maxima in the angular distributions for $\ell=0$, 2, 4, and 5 transfers, while the $\theta_{\rm lab}=18^{\circ}$ data point provided additional discrimination between the different $\ell$ transfers. These angles were chosen from the exploration of several DWBA calculations using the finite-range DWBA code Ptolemy~\cite{ptolemy}. Different global optical-model parametrizations for both deuterons~\cite{an,han,daehnick,bojowald} and $A=3$ ions~\cite{pang,liang,trost,hyakutake,becchetti} were explored. As has been observed in previous works at comparably high energies~\cite{kayge}, the angular distributions are less distinctive in shape than at energies nearer the Coulomb barrier.

Two different Faraday cups were used to integrate the beam current, depending on the angle of the GR spectrometer. At the most forward GR angle of $\theta_{\rm lab}=2.5^{\circ}$, the spectrometer aperture was obscured by the Faraday cup in the scattering chamber and so an alternative cup was used, located downstream of the scattering chamber. Several checks were made to ensure the two Faraday cups yielded consistent results. The transmission between the two Faraday cups was compared to a reference cup upstream in the beam line, which typically agreed at the 5\% level. Further, the Large Acceptance Spectrometer (LAS), also coupled to the scattering chamber with an aperture of 9~msr, was positioned at 60$^{\circ}$ throughout all measurements. This acted as a monitor detector for elastically scattered deuterons, independent of the choice of Faraday cup used for beam current integration. The LAS data were only used in longer runs where the statistics were sufficient; the typical count rate was of the order of $\sim$1 Hz. The fluctuations between the ratio of integrated beam current using different Faraday cups and the deuteron yield recorded in the LAS were less than 5\%.

\begin{figure}
\includegraphics[scale=0.7]{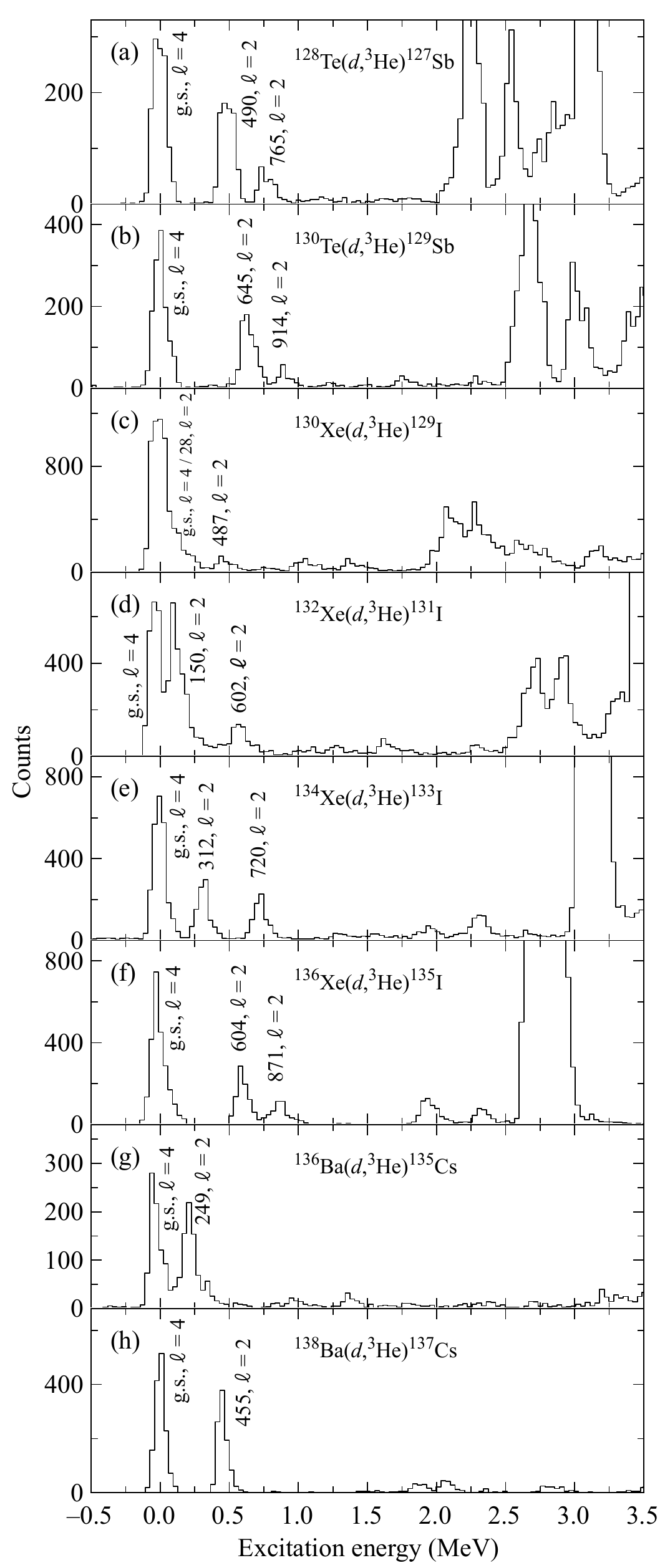}
\caption{\label{fig1} (a)--(h) Outgoing $^3$He spectra following the ($d$,$^3$He) reaction at an incident energy of 101~MeV on isotopes of $^{128,130}$Te, $^{130,132,134,136}$Xe, and $^{136,138}$Ba at $\theta_{\rm lab}=5.8^{\circ}$. The dominant peaks carrying proton strength corresponding to orbitals above $Z=50$ are labeled by their energy in keV and $\ell$ value.}
\end{figure}

\section{Analysis and Results} \label{analysis}

The outgoing $^3$He spectra are shown in Fig.~\ref{fig1} for the ($d$,$^3$He) reaction on $^{128,130}$Te, $^{130,132,134,136}$Xe, and $^{136,138}$Ba. The $Q$-value resolution was around 100~keV full width at half maximum, both for the solid and the Xe targets, and varied little over the angular range covered in these measurements. In all cases, excitation energy spectra were measured over a range of approximately 0-8~MeV; however, the states of interest are predominantly confined to the first 3~MeV in excitation energy. The states corresponding to excitations from below $Z=50$, initially with fragments of the $\pi0g_{9/2}$ strength, appear at excitation energies around 2-4~MeV. Strong peaks due to reactions on carbon and oxygen also appear in this region, and above. The characteristic features of the spectra below about 2~MeV in excitation energy include a 7/2$^+$ ground state, accounting for about half to three quarters of the proton occupancy above $Z=50$, followed by two weaker $\ell=2$ states, which in most cases appear to be of spin and parity 5/2$^+$, though some assignments of 3/2$^+$ have been made in the literature. This is referred to as $\ell=2$ or $\pi1d$ strength in the subsequent analysis. Common to all isotopes is that these first three states account for $\sim$80\% of the proton occupancy above $Z=50$. The remaining strength is shared between $2s_{1/2}$ and $0h_{11/2}$ proton orbitals, and some additional weak fragments of $1d$ and $0g_{7/2}$ strength. 

The cross sections were extracted from the yields, which were normalized to the integrated beam current and the product of the target-thickness and the aperture. Taking into account the sources of uncertainty discussed in Sec.~\ref{exp}, it is estimated that the systematic uncertainty on the absolute cross sections, dominated by the reliance on optical-model calculations, are $\lesssim$10\%. The systematic uncertainty on the relative cross sections, target to target, are estimated to be $\lesssim$6\%. The cross sections are tabulated in the Appendix. For cross sections larger than $\sim$50~$\mu$b/sr, the uncertainty is dominated by systematic uncertainty. Below that, the uncertainties are governed by statistics.

\subsection{DWBA and optical-model parameters }

\begin{figure}
\centering
\includegraphics[scale=0.7]{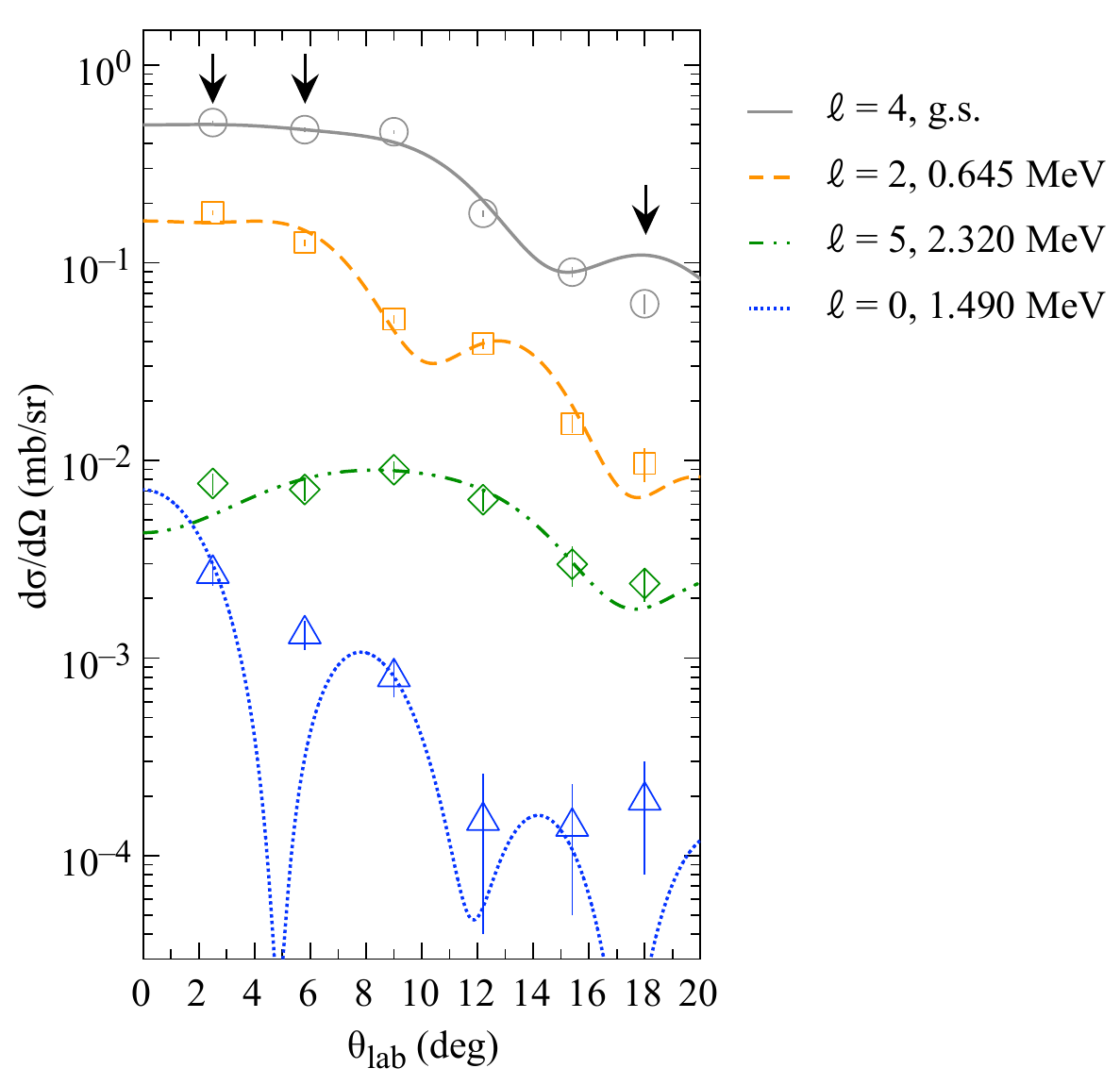}
\caption{\label{fig2} (color online). Angular distributions for the outgoing $^3$He ions following the $^{130}$Te($d$,$^3$He)$^{129}$Sb reaction at 101~MeV. The curves are DWBA calculations normalized to fit the data. Examples of $\ell=0$ (blue triangles, dotted line), 2 (orange squares, dashed), 4 (gray circles, solid), and 5 (green diamonds, dot-dashed) transfer are shown. Those for $\ell=0$, 2, and 5 are scaled by factors of 0.1, 0.5, and 0.2, respectively. The three arrows mark the angles at which measurements were made for the other targets. The error bars represent the statistical uncertainty only.}
\end{figure}

Figure~\ref{fig2} shows angular distributions for low-lying $\ell=0$, 2, 4, and 5 transitions in the $^{130}$Te($d$,$^3$He)$^{129}$Sb reaction, where cross sections were measured at six angles. Relatively good agreement is seen between the calculated angular distributions and the experimental data. In this case, the deuteron optical-model parameters of An and Cai~\cite{an} were used with those of Becchetti and Greenlees~\cite{becchetti} for $^3$He ions. Similar fits were achieved using the $^3$He optical-model potentials of Trost {\it et al.}~\cite{trost}. Poorer fits were obtained using $^3$He parametrizations of Refs~\cite{pang,hyakutake,liang}. Numerous deuteron parametrizations~\cite{han,daehnick,bojowald} were explored and little sensitivity was seen. The projectile wave function was given by the parametrizations of Brida {\it et al.}~\cite{brida}, based on Green's function Monte Carlo methods. The target bound-state wave function was generated using a Woods-Saxon potential with depth varied to reproduce the binding energy of the transferred nucleon; a radial parameter of $r_0=1.28$~fm, a diffuseness $a=0.76$~fm, and a spin-orbit potential characterized by $V_{\rm so}=6$~MeV, $r_{\rm so0}=1.09$~fm, and $a_{\rm so}=0.6$~fm were used.

With the high energy of the incident beam, there is good angular-momentum matching for high-$\ell$ transfer. For the first time, the $\ell=5$ strength was seen in each residual nucleus. For the $2s_{1/2}$ states that were seen, it is clear that at this high energy $\ell=0$ transfer is not well matched in angular momentum. However, there was good agreement with the DWBA-calculated angular distribution as shown in Fig.~\ref{fig2}. The 5.8$^{\circ}$ data lies close to a minimum and so is not a reliable angle to extract the spectroscopic strength; the 2.5$^{\circ}$ data were used to extract the $s$-state spectroscopic factors. 

\begin{table*}
\caption{\label{tab1} Proton occupancies deduced in this work.}
\newcommand\T{\rule{0pt}{3ex}}
\newcommand \B{\rule[-1.2ex]{0pt}{0pt}}
\begin{ruledtabular}
\begin{tabular}{lcccccc}
Isotope\B & $0g_{7/2}$ & $1d$ & $2s_{1/2}$ & $0h_{11/2}$ & Total & Expected \\
\hline
$^{128}$Te\T	&	1.13(9)	&	0.33(3)	&	0.012(10)	&	0.41(4)	&	1.87(10)	&	2	\\
$^{130}$Te	&	1.32(10)	&	0.32(3)	&	0.011(10)	&	0.24(3)	&	1.89(11)	&	2	\\
$^{130}$Xe	&	2.37(20)	&	1.00(11)	&	0.21(2)	&	0.37(3)	&	3.95(24)	&	4	\\
$^{132}$Xe	&	2.60(10)	&	0.94(5)	&	0.13(2)	&	0.41(4)	&	4.07(12)	&	4	\\
$^{134}$Xe	&	3.14(10)	&	0.71(4)	&	0.022(10)	&	0.37(4)	&	4.24(12)	&	4	\\
$^{136}$Xe	&	2.93(10)	&	0.52(3)	&	0.057(6)	&	0.40(4)	&	3.91(11)	&	4	\\
$^{136}$Ba	&	3.86(10)	&	1.29(8)	&	0.20(2)	&	0.62(6)	&	5.97(14)	&	6	\\
$^{138}$Ba\B	&	4.38(10)	&	1.15(8)	&	0.050(16)	&	0.59(7)	&	6.17(15)	&	6	\\
\hline
$^{130}$Xe$-^{130}$Te\T	&	1.05(23)	&	0.68(12)	&	0.20(2)	&	0.13(4)	&	2.06(26)	&	2	\\
$^{136}$Ba$-^{136}$Xe	&	0.93(14)	&	0.77(9)	&	0.14(2)	&	0.22(7)	&	2.06(18)	&	2	\\
\end{tabular}
\end{ruledtabular}
\end{table*}

A common normalization was used to determine the proton occupancies. For each isotope, $^{128,130}$Te, $^{130,132,134,136}$Xe, and $^{136,138}$Ba, the spectroscopic factor was extracted for each state populated. The results were summed and divided by the total proton occupancy expected above $Z=50$, namely 2 for the Te isotopes, 4 for Xe, and 6 for Ba. This produced eight independent normalization factors. The average value of all eight was used as a common normalization across all isotopes. Using the deuteron optical-model parametrizations of An and Cai~\cite{an} and $^3$He parametrizations of Becchetti and Greenlees~\cite{becchetti}, these were 0.566, 0.574, 0.598, 0.616, 0.642, 0.592, 0.603, and 0.623, yielding an average of 0.61 with an rms spread of 0.03 for the targets as listed above. Similar results were obtained for other optical-model parametrizations and are consistent with the typical values that one obtains from transfer reactions on stable isotopes~\cite{kay_quench}. This value, 0.61, was used for all the targets in the extraction of spectroscopic factors.

\subsection{Occupancies and uncertainties}
\label{specuncert}

\begin{figure*}
\centering
\includegraphics[scale=0.7]{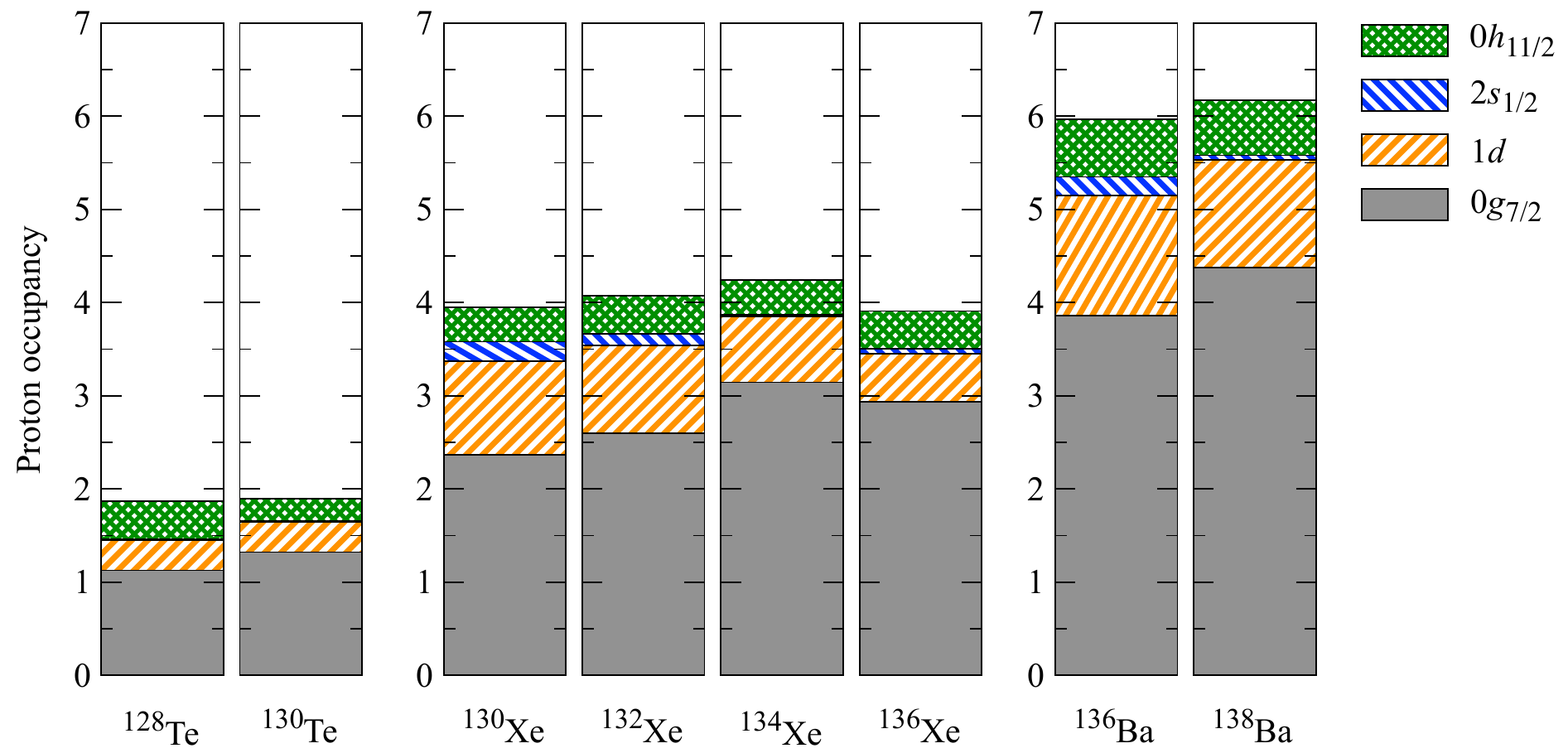}
\caption{\label{fig3} (color online). Ground-state proton occupancies beyond $Z=50$ for $^{128,130}$Te,  $^{130,132,134,136}$Xe, and $^{136,138}$Ba as derived from the experimentally determined cross sections. The uncertainties, discussed in the text, are estimated to be approximately $\pm$0.1 nucleons for each orbital.}
\end{figure*}

The summed valence proton occupancies above $Z=50$ are shown in Fig.~\ref{fig3} and Table~\ref{tab1} for the proton $0g_{7/2}$, $1d$, $2s_{1/2}$, and $0h_{11/2}$ orbitals. The dominant uncertainties are estimated to come from the spin assignments of weaker fragments, the spectroscopic factors for the 2$s_{1/2}$ strength, and from unassigned or mis-assigned strength. 

While $\sim$80\% of the proton strength lies in the first three strong states, numerous weak states carry the remaining strength. In many cases spin-parity assignments are available in the literature from $\beta$-decay studies and other $\gamma$-ray spectroscopy measurements and, though limited, from previous transfer-reaction experiments. In general, good agreement was found with existing assignments in the literature. In some cases it was not possible to make an assignment; this ``missed'' strength was small, less then a few percent in each case. This unassigned strength was not included in the sums to extract the normalization and thus contributes to the uncertainty. 

For $^{130}$Xe, an additional uncertainty arises from the ground-state doublet, comprising the 7/2$^+$ state at 0 keV and the 5/2$^+$ at 28 keV. This was fit as a doublet with the width fixed to that of an isolated state in the same spectrum, and the centroids constrained. A suspected doublet also occurs for the state around 2340~keV in excitation energy in $^{135}$I, which has a larger width than a single peak. It appears to be dominated by $\ell=5$ strength; the uncertainties for this strength are larger as a result.

To estimate the uncertainties from the optical-model parameters, the analysis was done with four different combinations of optical-model parametrizations and using different combinations of angles. The rms deviation on the summed strengths the four different analyses, carried out on all eight isotopes, was around 0.05-0.1 nucleons for each orbital. Further, using a single normalization, the total summed strengths are all within a few tenths of a nucleon, or $<$10\%, of the number of protons above $Z=50$, being 2, 4, and 6, for the Te, Xe, and Ba isotopes. It is difficult to state an uncertainty that can be applied to all the derived occupancies as there are some correlations in the extraction of the occupancies using different parametrizations and the common normalization procedure. Taking into account the evidence provided above, the uncertainty on the summed strength of any given orbital is estimated to be $\lesssim$0.1 nucleons. The uncertainties quoted in Table~\ref{tab1} reflect a combination of systematic and statistical uncertainties. For weak transitions, where multistep reactions become important, the spectroscopic factors have larger uncertainties (see, for example, Fig. 9 in Ref.~\cite{schifferni}). For transitions with cross sections weaker than 0.1~mb/sr, an additional uncertainty of $\pm$0.01 nucleons is added in quadrature. An additional $\pm$0.1 nucleons is added in quadrature to the uncertainties of the lowest lying $\ell=2$ and 4 strength in $^{130}$Xe due to the ground-state doublet.

\begin{figure*}[t]
\centering
\includegraphics[scale=0.75]{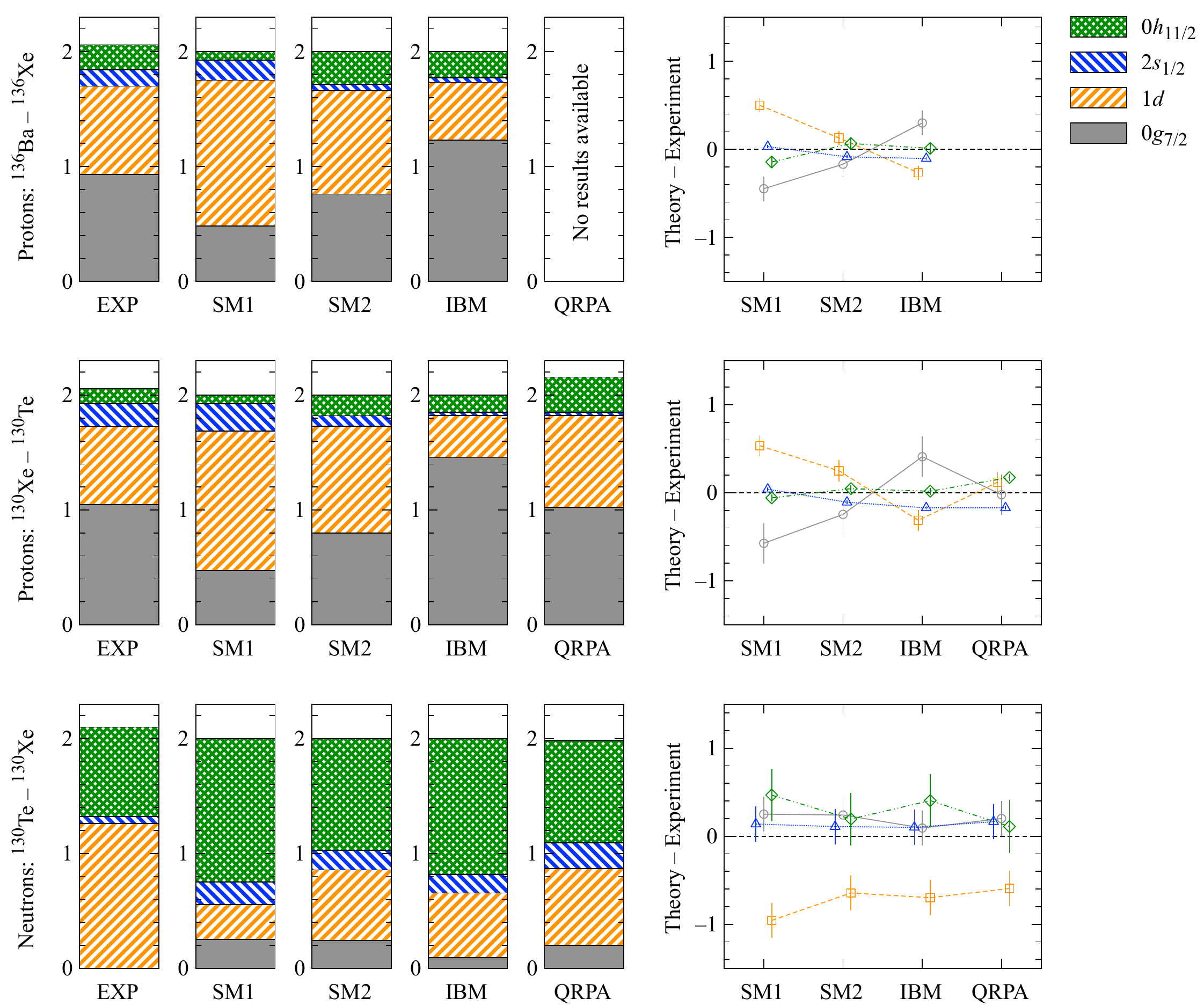}
\caption{\label{fig4}(color online). The bar charts to the left show the change in nucleon occupancies between the ground states for the $0\nu2\beta$-decay of $^{130}$Te~$\rightarrow$~$^{130}$Xe and $^{136}$Xe~$\rightarrow$~$^{136}$Ba.  The experimental data are denoted EXP. The proton data are from the current work, while the neutron data for the $^{130}$Te$\rightarrow^{130}$Xe system are from Ref.~\cite{kayxe}. The experimental data are compared to four different calculations:  SM1~\cite{neacsu}; SM2~\cite{menendez} (both shell-model calculations); IBM~\cite{kotila} (interacting-boson model); and QRPA~\cite{suhonen2010} (quasiparticle random-phase approximation). The plots to the right show a comparison of the theoretical calculations to the experimental data, for $2s_{1/2}$ (blue triangles, dotted line), $1d$ (orange squares, dashed), $0g_{7/2}$ (gray circles, solid), and $0h_{11/2}$ (green diamonds, dot-dashed) strength. The error bars reflect the uncertainty in the experimental data.}
\end{figure*}

\subsection{Comparison with other work}

There are few previous measurements with which to compare our results. The work of Auble {\it et al.}~\cite{auble_dh} reports on the ($d$,$^3$He) reaction at 34~MeV and Conjeaud {\it et al.}~\cite{conjeaud} on the ($t$,$\alpha$) reaction at 12~MeV, both on $^{128,130}$Te. Their results are in qualitative agreement with the current work in terms of the low-lying $\ell=2$ and 4 strength. Neither observed $\ell=5$ strength. Further, in the case of $\ell=4$ transfer to the ground state via ($d$,$^3$He) at 34~MeV, the cross sections were very small, around 50-100 $\mu$b/sr, suggesting that the angular-momentum matching was not ideal and that the analyses in both cases was done using local and zero-range DWBA calculations and with less refined global optical-model parametrizations.

\section{Discussion and theory}

The present results on proton occupancies, along with previous work probing the neutron vacancies~\cite{kayxe} of $^{130}$Te and $^{130}$Xe, completes a description of the ground-state valence nucleon occupancies for the $^{130}$Te~$\rightarrow$~$^{130}$Xe system. This allows us to quantitatively describe the {\it change} in neutron and proton occupancy in the $0\nu2\beta$-decay process. Any viable calculation of the nuclear matrix element should be also correctly describe these changes. 

Several theoretical calculations exist predicting both the neutron and proton occupancies of  $^{130}$Te, $^{130}$Xe, $^{136}$Xe, and $^{136}$Ba. Figure~\ref{fig4} shows a summary of experimental data and theoretical calculations describing the {\it change} in proton occupancies in the $0\nu2\beta$-decay process for the $^{130}$Te~$\rightarrow$~$^{130}$Xe and $^{136}$Xe~$\rightarrow$~$^{136}$Ba systems. Additionally, neutron vacancies from the experimental data from Ref.~\cite{kayxe} are also shown for the $^{130}$Te~$\rightarrow$~$^{130}$Xe system. The shell-model (SM) calculations are from Neacsu and Horoi (SM1)~\cite{neacsu} and from Men\'endez {\it et al.} (SM2)~\cite{menendez}. The quasiparticle random-phase approximation (QRPA) results refer to those denoted ``BCS+Adj.''\ in Suhonen and Civitarese~\cite{suhonen2010}. Results of a recent calculation using the interacting-boson model (IBM) by Kotila {\it et al.}~\cite{kotila} are shown also. The figure shows the difference between the theoretical calculations and the experimental data with the uncertainties in the experimental data included. This is to emphasize the discrepancies where present. These calculations were carried out before the experimental data was available, with the exception of the recent shell-model calculations (SM1) of Ref.~\cite{neacsu} and the IBM calculations of Ref.~\cite{kotila}, both of which were carried out after experimental data for the neutron vacancies were published, but before the current proton data were available.

\subsection{Proton occupancies} 

Focusing on the change in proton occupancies, we observe that the experimental changes between the parent and the daughter is mostly in the $\pi0g_{7/2}$ and $\pi1d$ orbitals, with the latter presumably being mostly the $\pi d_{5/2}$ strength. This is the same for both the $^{130}$Te~$\rightarrow$~$^{130}$Xe and $^{136}$Xe~$\rightarrow$~$^{136}$Ba decays, where the change in proton occupancies are, not surprisingly, similar. This is generally reflected in the calculations where there is, at least, a qualitative agreement. Both shell-model calculations, SM1 and SM2, overestimate the change in the $\pi1d$ orbital, with corresponding underestimate in the change of the $\pi0g_{7/2}$ orbital. The opposite is true of the IBM calculations. The SM2 results appear to provide a better description of the experimental data over the more recent SM1 calculations. For the $^{130}$Te~$\rightarrow$~$^{130}$Xe system, the QRPA calculations describe the change in proton occupancies very well. The plots highlight the fact that the calculations differ by $>$0.5 nucleons ($>$25\%) in some cases, and most importantly this is in the cases of the $\pi 0g_{7/2}$ and $\pi 1d$ orbitals, which are dominant in this process and likely have significant impact on the magnitude of the nuclear matrix element. Within the experimental uncertainties, it is difficult to draw conclusions about the $\pi2s_{1/2}$ and $\pi0h_{11/2}$ strengths; they play only small roles in the occupancy and thus one would not expect them to play a major role in the magnitude of the nuclear matrix element. 

\subsection{Neutron vacancies}

The valence neutrons participating in the decay are just below $N=82$. While the valence orbital space is the same for neutrons as that for protons, it is found to be truncated, with the $\nu0g_{7/2}$ orbital playing no observable role in the change between the $^{130}$Te and $^{130}$Xe ground states. An experimental limit of $<$0.1 nucleons in the vacancy of the $\nu0g_{7/2}$ orbital, set in the neutron-adding ($\alpha$,$^3$He) reaction at 50 MeV which are conditions favorable for the population of $\ell=4$ strength, has been made in Ref.~\cite{kayxe}. Aside from this, the most noticeable feature in the comparison between theory and experiment is the significant underestimation of the change in $\nu1d$ strength, assumed to be predominantly the $\nu d_{3/2}$ strength, in the calculations. There appears to be quite good agreement for the other orbitals, though this agreement is perhaps augmented by the lack of $\nu0g_{7/2}$ in the experimental data.

\subsection{General comments}

Any calculations used to determine nuclear matrix elements should be able to reproduce the nucleon occupancies, and how they change in the decay process, to a reasonable degree of precision; at present they do not. The experimental data, within uncertainties, reflect the change in the 0$^+$ ground-state wave functions. The occupancy of the valence orbitals is one nuclear-structure property that may help constrain the nuclear matrix element calculations. Other features of the nuclear structure are being explored along with alternative approaches to calculating nuclear-matrix-element calculations. An important feature probed in two-nucleon transfer reactions is the presence of pairing vibrations. These are characterized by strong 0$^+$ excitations which represent a second sea of correlated neutrons or protons. These are likely to complicate calculations, particular those such as QRPA, which cannot account for such features. Such pairing vibrations have been observed for protons in the the proximity of $^{130}$Te, $^{130}$Xe~\cite{alford_te}, and indeed $^{136}$Xe and $^{136}$Ba~\cite{alford_other}. They are not present for neutrons~\cite{kayxe}. Hybrid models have been considered~\cite{bes} in which neutrons are treated in a superfluid phase and protons in a normal phase. Data from two-nucleon transfer has been discussed in other contexts too. For example, Ref.~\cite{brown} discusses the nuclear matrix elements in terms of an expansion over states in the $A-2$ systems, such that data from the ($p$,$t$) and ($^3$He,$n$) reactions may become important in this context. For the present system, this would connect $^{130}$Te to $^{130}$Xe via $^{130}$Te$(p,t)^{128}$Te($^3$He,$n$)$^{130}$Xe. In this case, data for both of these reactions exist~\cite{kayxe,alford_te}. 

\section{Conclusion}

We report on the determination of proton occupancies from data on the ($d$,$^3$He) reaction on isotopes of $^{128,130}$Te, $^{130,132,134,136}$Xe, and $^{136,138}$Ba. This work has provided a quantitative description of the {\it change} in the proton occupancies between the 0$^+$ ground states on the $0\nu2\beta$-decay candidates, $^{130}$Te~$\rightarrow$~$^{130}$Xe and $^{136}$Xe~$\rightarrow$~$^{136}$Ba, and complements recent data mapping out the neutron vacancies of the  $^{130}$Te~$\rightarrow$~$^{130}$Xe system. There is a quantitative disagreement between the experimental data and recent calculations of the same properties. There is no particular model, at least from comparisons with the results of SM, IBM, and QRPA calculations, that fully describes the experimental occupancy data, and therefore the nuclear structure of the isotopes involved in $0\nu2\beta$-decay, better than the others. It is hoped that these data provide an important constraint on future calculations of the nuclear matrix elements.

\section {Acknowledgements}

This measurement (Experiment E399) was performed at Research Center for Nuclear Physics at Osaka University. The authors wish to thank the RCNP operating staff. Participants from the U.S. and U.K. would like to express their gratitude to the local staff and administration for their hospitality and assistance. We would like to thank John Greene for preparing targets for this experiment and Brad DiGiovine for the handling of the gaseous isotopes. This material is based upon work supported by the U.S. Department of Energy, Office of Science, Office of Nuclear Physics, under Contract No. DE-AC02-06CH11357, and by the UK Science and Technology Facilities Council.

\appendix*
\section{}
\label{appendix}

Cross sections for the ($d$,$^3$He) reaction on $^{128,130}$Te, $^{130,132,134,136}$Xe, and $^{136,138}$Ba are given in Tables~\ref{tab2}--\ref{tab9} along with normalized spectroscopic factors. The energies and spin-parity assignments are taken from the literature~\cite{nndc}, where known. States with tentative spin-parity assignments that are newly observed in this work, are shown in parentheses. We adopt a tentative assignment if that is also what appears in the database~\cite{nndc}. Where the energy of a state is provided from the current analysis it is rounded to the nearest 10 keV, reflecting the estimated uncertainty of $\pm$10~keV. The uncertainties on cross sections below $\sim$50~$\mu$b/sr are dominated by statistical uncertainties, becoming larger than about 5-10\%. For cross sections larger than that, the systematic uncertainties are the dominant uncertainty. The systematic uncertainties on the absolute magnitude of the cross sections are estimated to be  around 10\% (discussed in Sec.~\ref{analysis}) due to the determination of the target thickness using high-energy deuteron scattering, and thus relying on model-dependent optical-model parametrizations. The systematic uncertainties on the relative cross sections are estimated to be of the order of 6\%. Uncertainties on the normalized spectroscopic factors follow the prescription laid out in Sec.~\ref{specuncert}. The normalization is achieved such that the total occupancies for the relevant orbits add up tot he number of protons beyond $Z=50$. This normalization factor is a single number, 0.61, which represents the average over all eight targets, and is independent of the target mass and $\ell$ value.

\begin{table}[]
\caption{\label{tab2} Experimental differential cross sections ($\sigma$) in $\mu$b/sr for $^{128}$Te($d$,$^3$He)$^{127}$Sb reaction at 101~MeV are given for $\theta_{\rm lab}=2.5^{\circ}$, 5.8$^{\circ}$, and 18$^{\circ}$. Normalized spectroscopic factors are also given. Those with $\ell$ values in parentheses have tentative assignments, either from this work or in the literature~\cite{nndc}. Energies are in keV, with values from the literature where known. Uncertainties are discussed in the text.}
\newcommand\T{\rule{0pt}{3ex}}
\newcommand \B{\rule[-1.2ex]{0pt}{0pt}}
\begin{ruledtabular}
\begin{tabular}{lccccc}
$E$\B & $\ell$ & $\sigma_{2.5^{\circ}}$ & $\sigma_{5.8^{\circ}}$ & $\sigma_{18^{\circ}}$ &$C^2S$ \\
\hline
0\T	&	4	&	510	&	419	&	54	&	1.13	\\
491	&	2	&	375	&	280	&	19	&	0.23	\\
765	&	2	&	140	&	84	&	8.8	&	0.07	\\
1050	&	--	&	9.1	&	8.5	&	0.9	&	--	\\
1186	&	0	&	29	&	16	&	1.8	&	0.01	\\
1352	&	(2)    &	9.5	&	6.0	&	--	&	$<$0.01	\\
1610	&	(2)	&	23	&	14	&	3.6	&	0.01	\\
1790&	(2)	&	23	&	16	&	1.8	&	0.01	\\
1950	&	--	&	8.0	&	7.8	&	1.8	&	--	\\
2130\footnote{State lies close to the previously 2124- and 2140-keV states, both reporting possible 11/2$^-$ assignments.}	&	(5)	&	95	&	89	&	19	&	0.41	\\

\end{tabular}
\end{ruledtabular}
\end{table}

\begin{table*}[]
\caption{\label{tab3} $^{130}$Te($d$,$^3$He)$^{129}$Sb (notation same as Table~\ref{tab2}). Cross sections for additional angles of $\theta_{\rm lab}=9.0^{\circ}$, 12.2$^{\circ}$, and 15.4$^{\circ}$ are given.}
\newcommand\T{\rule{0pt}{3ex}}
\newcommand \B{\rule[-1.2ex]{0pt}{0pt}}
\begin{ruledtabular}
\begin{tabular}{lcccccccc}
$E$\B & $\ell$ & $\sigma_{2.5^{\circ}}$ & $\sigma_{5.8^{\circ}}$ & $\sigma_{9.0^{\circ}}$ & $\sigma_{12.2^{\circ}}$ & $\sigma_{15.4^{\circ}}$ & $\sigma_{18^{\circ}}$ &$C^2S$ \\
\hline
0\T	&	4	&	471	&	453	&	441	&	165	&	85	&	61	&	1.32	\\
645	&	2	&	329	&	243	&	99	&	72	&	29	&	19	&	0.21	\\
914	&	2	&	67	&	55	&	17	&	14	&	5.7	&	4.4	&	0.05	\\
1220	&	(2)	&	35	&	20	&	9.6	&	4.6	&	1.8	&	3.7	&	0.02	\\
1493	&	(0)	&	25	&	13	&	7.6	&	1.4	&	1.3	&	1.8	&	0.01	\\
1762	&	(2)	&	58	&	40	&	18	&	13	&	4.8	&	3.3	&	0.04	\\
2020\footnote{Close in energy to the previously reported 2031-keV state, which included a tentative 11/2$^-$ assignment.}	&	(5)	&	20	&	13	&	17	&	10	&	2.2	&	1.5	&	0.06	\\
2150	&	(2)	&	16	&	13	&	6.1	&	3.2	&	5.3	&	4.0	&	0.01	\\
2320\footnote{Possibly corresponds to the previously observed 2317-keV state that has $j=11/2$ as a possible spin assignment.}	&	(5)	&	35	&	34	&	43	&	29	&	14	&	12	&	0.18	\\
\end{tabular}
\end{ruledtabular}
\end{table*}

\begin{table}[h]
\caption{\label{tab4} $^{130}$Xe($d$,$^3$He)$^{129}$I (notation same as Table~\ref{tab2}).}
\newcommand\T{\rule{0pt}{3ex}}
\newcommand \B{\rule[-1.2ex]{0pt}{0pt}}
\begin{ruledtabular}
\begin{tabular}{lccccc}
$E$\B & $\ell$ & $\sigma_{2.5^{\circ}}$ & $\sigma_{5.8^{\circ}}$ & $\sigma_{18^{\circ}}$ &$C^2S$ \\
\hline
0\T	&	4	&	1061	&	966	&	140	&	2.37	\\
28	&	2	&	782	&	874	&	84	&	0.71	\\
278	&	2	&	156	&	61	&	6.9	&	0.05	\\
487	&	2	&	158	&	141	&	13	&	0.11	\\
560	&	0	&	24	&	15	&	7.3	&	0.01	\\
1047	&	2	&	212	&	137	&	16	&	0.11	\\
1230	&	--	&	41	&	31	&	--	&	--	\\
1401\footnote{Tentatively assigned $\ell=5$, 9/2$^-$ in a previous measurement, though highly likely it is $\ell=5$, 11/2$^-$.}	&	(5)	&	111	&	106	&	30	&	0.37	\\
1566	&	(0)	&	49	&	20	&	6.3	&	0.02	\\
1741\footnote{A possible $\ell=0+4$ doublet reported in previous work. Cannot assign in the present work.}	&	--	&	40	&	14	&	1.3	&	--	\\
1861	&	2	&	41	&	24	&	2.8	&	0.02	\\
2012	&	0	&	400	&	160	&	36	&	0.18	\\
\end{tabular}
\end{ruledtabular}
\end{table}

\begin{table}[h]
\caption{\label{tab5} $^{132}$Xe($d$,$^3$He)$^{131}$I (notation same as Table~\ref{tab2}).}
\newcommand\T{\rule{0pt}{3ex}}
\newcommand \B{\rule[-1.2ex]{0pt}{0pt}}
\begin{ruledtabular}
\begin{tabular}{lccccc}
$E$\B & $\ell$ & $\sigma_{2.5^{\circ}}$ & $\sigma_{5.8^{\circ}}$ & $\sigma_{18^{\circ}}$ &$C^2S$ \\
\hline
0\T	&	4	&	948	&	976	&	127	&	2.60	\\
150	&	2	&	711	&	723	&	67	&	0.62	\\
493	&	2	&	94	&	68	&	10	&	0.06	\\
602	&	2	&	201	&	202	&	5.2	&	0.17	\\
877	&	0	&	42	&	33	&	6.5	&	0.02	\\
1020	&	--	&	24	&	36	&	12	&	--	\\
1147	&	2	&	41	&	40	&	--	&	0.03	\\
1298	&	2	&	79	&	66	&	3.9	&	0.06	\\
1435	&	--	&	14	&	22	&	5.2	&	--	\\
1646	&	5	&	87	&	101	&	22	&	0.41	\\
1718	&	0	&	74	&	18	&	6.5	&	0.03	\\
1860&	--	&	24	&	24	&	7.8	&	--	\\
2020	&	--	&	13	&	18	&	2.6	&	--	\\
2130	&	--	&	17	&	13	&	--	&	--	\\
2308	&	0	&	151	&	99	&	23	&	0.07	\\
\end{tabular}
\end{ruledtabular}
\end{table}

\begin{table}[h]
\caption{\label{tab6} $^{134}$Xe($d$,$^3$He)$^{133}$I (notation same as Table~\ref{tab2}).}
\newcommand\T{\rule{0pt}{3ex}}
\newcommand \B{\rule[-1.2ex]{0pt}{0pt}}
\begin{ruledtabular}
\begin{tabular}{lccccc}
$E$\B & $\ell$ & $\sigma_{2.5^{\circ}}$ & $\sigma_{5.8^{\circ}}$ & $\sigma_{18^{\circ}}$ &$C^2S$ \\
\hline
0\T	&	4	&	914	&	1094	&	162	&	3.14	\\
312	&	2	&	435	&	417	&	31	&	0.37	\\
720	&	2	&	330	&	305	&	29	&	0.28	\\
945	&	--	&	13	&	20	&	7.5	&	--	\\
1313	&	2	&	43	&	39	&	2.5	&	0.04	\\
1455	&	--	&	7.3	&	13	&	1.7	&	--	\\
1564	&	0	&	44	&	33	&	3.4	&	0.02	\\
1730	&	--	&	28	&	14	&	7.5	&	--	\\
1910	  &	--	&	60	&	36	&	11	&	--	\\
1980\footnote{Close in the energy to the 1991-keV previously reported in the literature, which has a possible $11/2^-$ spin-parity assignment.}	&	5	&	40	&	79	&	6.7	&	0.37	\\
2150	&	--	&	--	&	27	&	5.9	&	--	\\
2467	&	--	&	33	&	21	&	3.4	&	--	\\
2580	&	--	&	19	&	16	&	--	&	--	\\
2680	&	--	&	129	&	52	&	4.2	&	--	\\
2825	&	(2)	&	23	&	21	&	1.7	&	0.02	\\

\end{tabular}
\end{ruledtabular}
\end{table}

\begin{table}[h]
\caption{\label{tab7} $^{136}$Xe($d$,$^3$He)$^{135}$I (notation same as Table~\ref{tab2}).}
\newcommand\T{\rule{0pt}{3ex}}
\newcommand \B{\rule[-1.2ex]{0pt}{0pt}}
\begin{ruledtabular}
\begin{tabular}{lccccc}
$E$\B & $\ell$ & $\sigma_{2.5^{\circ}}$ & $\sigma_{5.8^{\circ}}$ & $\sigma_{18^{\circ}}$ &$C^2S$ \\
\hline
0\T	&	4	&	920	&	972	&	150	&	2.93	\\
604	&	2	&	398	&	375	&	28	&	0.35	\\
871	&	2	&	203	&	185	&	12	&	0.17	\\
1370	&	--	&	11	&	12	&	--	&	--	\\
2340\footnote{Peak suspected to be a doublet due to larger width, perhaps with the 2312-keV state reported in the literature with unassigned spin-parity.}	&	(5)	&	77	&	75	&	29	&	0.40	\\
3110	&	(0)	&	113	&	73	&	5.2	&	0.06	\\
3320	&	--	&	36	&	23	&	7.2	&	--	\\
3620	&	--	&	16	&	9.3	&	6.4	&	--	\\
\end{tabular}
\end{ruledtabular}
\end{table}

\begin{table}[h]
\caption{\label{tab8} $^{136}$Ba($d$,$^3$He)$^{135}$Cs (notation same as Table~\ref{tab2}).}
\newcommand\T{\rule{0pt}{3ex}}
\newcommand \B{\rule[-1.2ex]{0pt}{0pt}}
\begin{ruledtabular}
\begin{tabular}{lccccc}
$E$\B & $\ell$ & $\sigma_{2.5^{\circ}}$ & $\sigma_{5.8^{\circ}}$ & $\sigma_{18^{\circ}}$ &$C^2S$ \\
\hline
0\T	&	4	&	1301	&	1394	&	174	&	3.65	\\
250\footnote{Includes a possible contribution to the yield of less than a few percent of the total due to an isotopic contaminant.}	&	2	&	1300	&	1071	&	71	&	0.96	\\
408	&	(2)	&	175	&	87	&	5.2	&	0.08	\\
608	&	2	&	75	&	56	&	7.0	&	0.05	\\
790\footnote{Appears at the same energy as a state previously reported at $11/2^+$ in the literature. Includes a possible contribution to the yield of $\lesssim$50\% of the total due to an isotopic contaminant.}	&	(2)	&	77	&	61	&	3.1	&	0.06	\\
1030	&	(0)	&	187	&	120	&	20	&	0.10	\\
1150	&	--	&	83	&	29	&	7.0	&	--	\\
1420	&	(5)	&	162	&	169	&	37	&	0.62	\\
1690	&	(0)	&	123	&	41	&	3.9	&	0.07	\\
1880	&	(2)	&	64	&	58	&	4.4	&	0.05	\\
2140	&	--	&	66	&	42	&	2.6	&	--	\\
2310	&	(4)	&	64	&	59	&	10	&	0.22	\\
2470	&      (2)	 &	164	&	96	&	5.7	&	0.09	\\
2620	&	(0)	&	57	&	18	&	2.6	&	0.03	\\
2770	&	--	&	84	&	59	&	7.0	&	--	\\
2930	&	--	&	53	&	29	&	--	&	--	\\

\end{tabular}
\end{ruledtabular}
\end{table}

\begin{table}[h]
\caption{\label{tab9} $^{138}$Ba($d$,$^3$He)$^{137}$Cs (notation same as Table~\ref{tab2}).}
\newcommand\T{\rule{0pt}{3ex}}
\newcommand \B{\rule[-1.2ex]{0pt}{0pt}}
\begin{ruledtabular}
\begin{tabular}{lccccc}
$E$\B & $\ell$ & $\sigma_{2.5^{\circ}}$ & $\sigma_{5.8^{\circ}}$ & $\sigma_{18^{\circ}}$ &$C^2S$ \\
\hline
0\T	&	4	&	1431	&	1556	&	395	&	4.38	\\
455	&	2	&	1137	&	1076	&	155	&	1.01	\\
830	&	--	&	11	&	12	&	6.1	&	--	\\
1120 &	--	&	14	&	10	&	--	&	--	\\
1490	&	0	&	19	&	21	&	1.8	&	0.01	\\
1620	&	--	&	20	&	19	&	0.9	&	--	\\
1868\footnote{Assigned 9/2$^-$ in the literature, but most likely 11/2$^-$ due to its strength in this reaction.}	&	5	&	122	&	137	&	67	&	0.59	\\
2068	&	2	&	169	&	144	&	33	&	0.14	\\
2150	&	0	&	71	&	36	&	--	&	0.04	\\
2350	&	--	&	23	&	28	&	11	&	--	\\
2520	&	--	&	15	&	16	&	11	&	--	\\
2796&	--	&	143	&	93	&	7.9	&	--	\\
2910	&	--	&	108	&	57	&	18	&	--	\\
3020	&	--	&	35	&	25	&	10	&	--	\\
3190	&	--	&	31	&	4.4	&	--	&	--	\\

\end{tabular}
\end{ruledtabular}
\end{table}


\end{document}